# DEVELOPMENT OF AI-CLOUD BASED HIGH-SENSITIVITY WIRELESS SMART SENSOR FOR PORT STRUCTURE MONITORING

*Junsik Shin, Junyoung Park & Jongwoong Park*

*Chung-Ang University, Korea*

**ABSTRACT:** *Regular structural monitoring of port structure is crucial to cope with rapid degeneration owing to its exposure to saline and collisional environment. However, most of the inspections are being done visually by human in irregular-basis. To overcome the complication, lots of research related to vibration-based monitoring system with sensor has been devised. Nonetheless, it was difficult to measure ambient vibration due to port's diminutive amplitude and specify the exact timing of berthing, which is the major excitation source. This study developed a novel cloud-AI based wireless sensor system with high-sensitivity accelerometer EPSON M-A352, which has 0.2µG/√Hz noise density, 0.003mg of ultra-low noise feature, and 1000Hz of sampling frequency. The sensor is triggered based on either predefined schedule or long rangefinder. After that, the detection of ship is done by AI object detection technique called Faster R-CNN with backbone network of ResNet for the convolution part. Coordinate and size of the detected anchor box is further processed to certify the berthing ship. Collected data are automatically sent to the cloud server through LTE CAT 1 modem within 10Mbps. The system was installed in the actual port field in Korea for few days as a preliminary investigation of proposed system. Additionally, acceleration, slope, and temperature data are analyzed to suggest the possibility of vibration-based port condition assessment.*

## 1. INTRODUCTION

In the field of civil engineering, research of adequate monitoring system accompanied with state-of-the-art technology for the infrastructure condition assessment has shown great expansion owing to arising degradation and ageing issues constantly being reported. Port structure is also no exception. It is even more vulnerable than typical inland structures due to its environmental uniqueness. Exposure to the coastal environment with saline substances stimulates degradation of the structure which is mainly composed of concrete and steel. In addition, periodic sinusoidal impact generated by the waves and berthing excitation from the large ship may weaken the structure. Moreover, rampant climate change results in frequent heavy rainfall and typhoons, which makes the port structure to barely sustain throughout its lifespan.

For preemptive condition assessment of the port structure, visual inspection has been implemented from the past as a basic maneuver. The visual inspection can clearly verify the cracks and external defects on the surface, while it is irregular, time consuming, and subjective to the investigator's personal view. In addition, accessing directly to the underwater for the inspection can be dangerous and often hindered by the murky sight and shells stuck on the surface. Choi et al. (2017) suggested robot based underwater inspection system to replace human, but this kind of visual inspection has a chronic issue; it is infeasible to detect the inherent damage and structural stability which are the key components of accurate condition assessment.

To overcome the drawbacks of visual inspection, vibration-based structural monitoring system has been addressed with the help of technological advancement in Micro-Electro Mechanical Systems (MEMS) and various types of sensors. Zarafshan et al. (2012) proposed bridge scour monitoring system by analyzing time and frequency domain of vibration data acquired from the pier. The research has successfully verified the possibility of structural health monitoring through vibration analysis. Unfortunately, the related research has mostly targeted inland structures like buildings and bridges, while coastal structures are slackened off due to several reasons: 1) Vibration level of port structure having large mass is diminutive. Min et al. investigated constant vibration sensing system with ADXL354 MEMS accelerometer and applied to a wharf in Korea. However, it was figured out that the collected data did barely contain vibrational information as the actual amplitude of the structural vibration was way lower than the sensor's sensitivity (Min, 2022). 2) Constant sensing system was not adequate for port structure in aspects of accessibility to the site and data processing. Park et al. (2011) proposed Imote2 sensor to the port monitoring, however, this kind of system may not last for longer period owing to the constant sensing system which consumes lots of energy. For the port structure, a long-term monitoring is important as the port is usually hard to access and its critical structure change is unlikely to occur in short term aspect. Besides, overflow of data with irrelevant information may waste the storage of the server and even require human intervention to parse and extract data. Due to these kinds of technological difficulties, vibration-based port monitoring could not be assessed properly.

The study herein focuses on the development of AI-Cloud based monitoring system that accommodates overall limitations stated beforehand. First, development of the sensing system is introduced. This part covers contents about various types of sensors used for data processing and acquisition such as Micro-Controller Unit (MCU), power supply system, accelerometer, and wireless communication modem. Next, applied trigger and AI-Cloud systems are addressed. This section describes about one of the most differentiated features compared to the ordinary systems, which is the identification of the ship berthing. This feature encourages the system to be operated by the event, hence it can capture the major excitation efficiently for vibration-based condition assessment and save the energy. The suggested cloud system is open-ended and real-time, which makes it possible to efficiently add the additional sensor nodes and provide direct feedback to the site for the proper action. Finally, the deployment of the system to the field in short-term and the vibration data automatically analyzed through the cloud system are discussed further for the actual condition assessment.

## 2. DEVELOPMENT OF PORT MONITORING SYSTEM

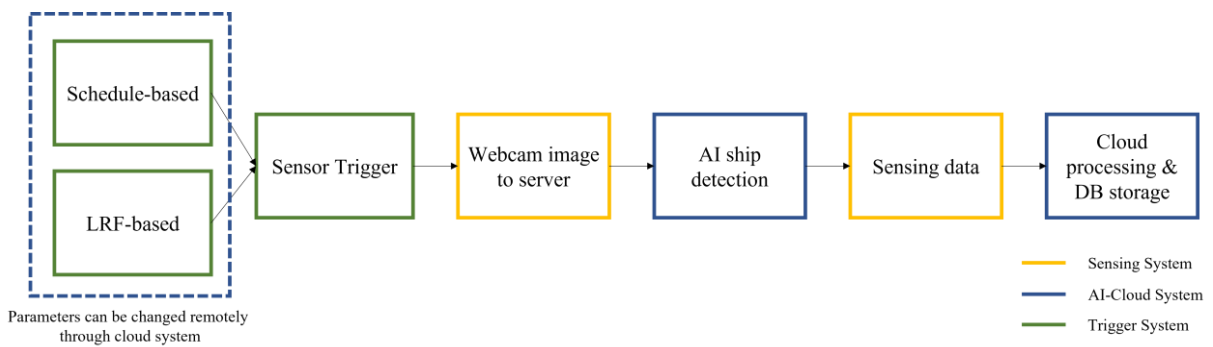

Fig. 1: Sensor process algorithm diagram

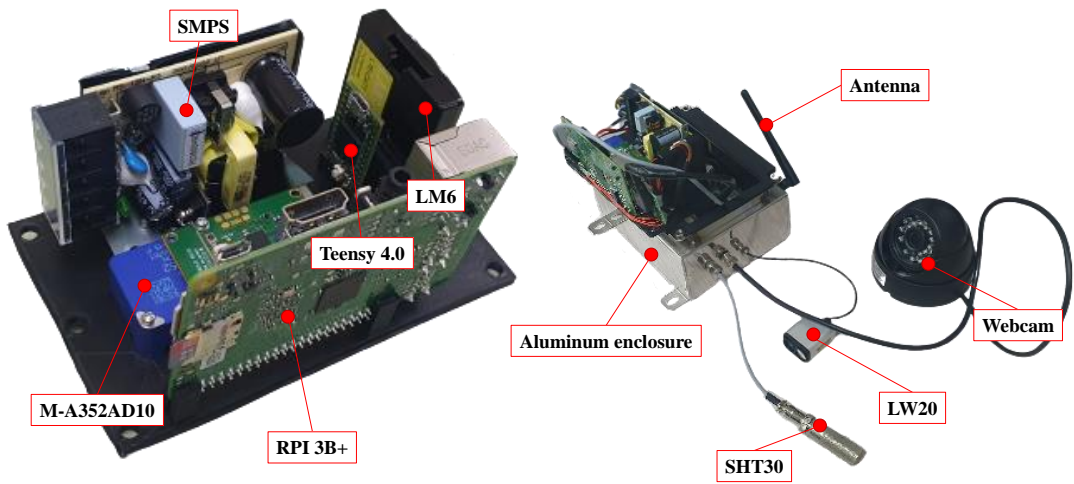

Fig. 2: Overview of the proposed system's hardware

To enable sustainable smart health monitoring system of the port structure, a specialized wireless sensor system was designed. First part on the Figure 1 is the trigger system which initiates overall system cycle. The trigger system consists of two major routes, which are schedule-based and long rangefinder (LRF) based. According to the predefined logics and trigger system, sensing system gets triggered. The sensing system does not get started all the time to save the power and alleviate acquisition of unnecessary data. First step of sensing system is capturing of image through connected webcam. The captured image is then directly sent to the cloud server via LTE communication for ship detection. Embedded AI object detection technique called Faster R-CNN trained with custom ship dataset is processed to detect the presence of the ship within the image promptly (Ren, S. et al., 2017). Depending on the prediction, sensing time is determined. If ship is not recognized, short sensing time is applied for sensor status checkup and further long-term trend analysis. If ship is recognized, on the other hand, longer sensing time is assigned to exactly measure vibration throughout the overall berthing process. On the sensing system, physical quantities such as acceleration, tilt, distance, and temperature are measured. Collected data are sent to the server wirelessly and vibration analysis is automatically processed. After the cycle is finished, the sensor goes back to the rest mode and waits for the next trigger.

## 2.1 Sensing System

### 2.1.1 MCU and Power supply

MCU works as a core of the sensor system by commanding all the connected sensors and processing data retrieved. Therefore, it is important to select an adequate MCU that satisfies the purpose of the system. For the proposed system, Raspberry Pi (RPI) 3B+ with Linux OS has been applied as a main MCU. Selected MCU has 1.4GHz a quad-core processor, which is suitable for high frequency data acquisition and computation process without any latency. It also provides multiple USB ports and various types of communication protocols such as UART, SPI, and I2C, which are suitable for proposed sensing system. A most recent model, RPI 4, has also been considered, but outdoor application might aggravate its noted overheat issues which would lead to failure of the total system.

The power supply of the system was designed to have two different options: constant power mode and solar charging mode. If constant power source is affordable in the site, connected AC power can be converted to 5V DC power through a switched mode power supply (SMPS) which can power up the MCU. If constant power outage occurs or is unaffordable on the site, second option can be chosen. Embedded rechargeable lithium-ion batteries are automatically used as a subordinate expedient. The batteries are charged with the solar panel and LT3652 charging integrated circuit. Moreover, proposed cloud computing and trigger techniques mitigate power consumption drastically, which leads the system to be operated with low power.

### 2.1.2 EPSON M-A352AD10 Accelerometer

One of the factors that limited vibration-based port monitoring is port's diminutive amplitude of vibration. Therefore, accelerometer having high sensitivity and resolution was selected prudently. Selected sensor is EPSON's M-A352AD10 accelerometer, featuring digital tri-axial acceleration output, ultra-low noise, high stability, and low power consumption. Its ultra-low noise feature provides $0.2\mu G/\sqrt{Hz}$ of noise density, which is technically difficult to be achieved for typical MEMS accelerometer. For the sensitivity, its scale factor is $2^{-24}$G/LSB, showing that it can sense any subtle vibration. M-A352AD10 is connected with the MCU with USB Type-C for communication. To start sensing, MCU commands the accelerometer by Python script, and depending on the trigger type (Section 2.2), sensing time determined. Furthermore, reduced noise mode and temperature shock stabilization mode are added to make the sensor to be tenacious against external factors.

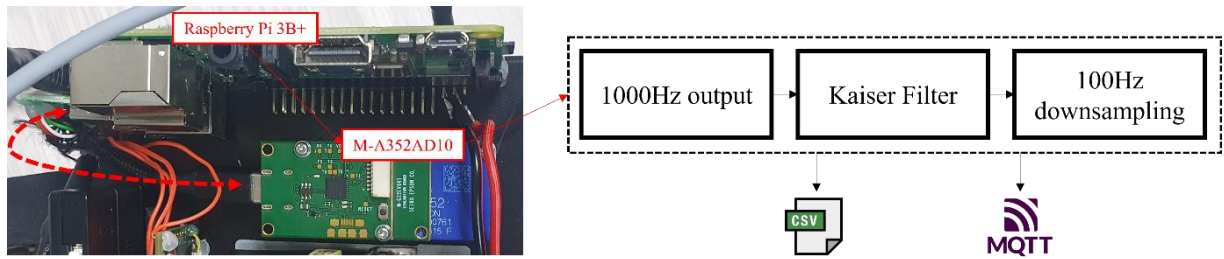

Fig. 3: Acceleration acquisition process

As soon as the sensing starts by the command from the MCU, the accelerometer starts to burst out tri-axial acceleration data in 1000Hz of sampling frequency. While sampling, digital Finite Impulse Response (FIR) low pass filter called Kaiser Filter with 128 Taps and cutoff frequency of 110Hz is applied to remove high frequency noise. After that, data are downsampled to 100Hz in real time. While sensing, data are directly saved into CSV file and sent to server through Message Queuing Telemetry Transport (MQTT) protocol at the same time. (Section 2.1.5)

### 2.1.3 Tilt estimation from acceleration

Tilt of the structure can provide useful information about structure's stability in long-term aspect. Most of the tilt sensors are based on the biaxial or triaxial accelerometer and changes in acceleration is used to estimate the rotation and tilt. Therefore, it is beneficial to calculate tilt while doing acceleration measurement. Tilt estimation from

acceleration can be expressed as equation (1.1):

$$\theta = \arctan\left(\frac{a_x}{\sqrt{a_y^2 + a_z^2}}\right), \psi = \arctan\left(\frac{a_y}{\sqrt{a_x^2 + a_z^2}}\right) \tag{1.1}$$

where $\theta, \psi$ denotes pitch and roll tilt angle respectively; $a_x, a_y, a_y$ denotes x, y, z axis acceleration outputted from accelerometer respectively. While accelerometer can estimate pseudo-static component of tilt successfully, dynamic component of tilt is limited due to translational acceleration such as structural vibration and signal noise generated by the sensor crosstalk (Liu et al., 2017). To confront the limitation, Liu et al. proposed a complementary filter to take advantage of both accelerometer which can obtain pseudo-static component and gyroscope which is accurate for dynamic component. Liu has also experimentally proved that tilt with dynamic range of 0.001-0.1Hz can be accurately estimated by accelerometer. As a result, proposed sensing system has implemented real-time low pass filter (LPF) with cutoff frequency of 1Hz to the tilt for accurate data acquisition.

### 2.1.4  Validation of acceleration and tilt

To experimentally validate performance of the proposed accelerometer, noise floor comparison with typical MEMS accelerometer ADXL354 was conducted. Both accelerometers were placed together on less-vibration place and the data were synchronously collected for one minute with sampling rate of 100Hz. The root mean square error (RMSE) value of ADXL354's measured noise was 0.1mg, while M-A352AD10's was 0.003mg. Through the experiment, it was verified that proposed sensing system is able to sense diminutive level of vibration with around 300 times better than the typical MEMS accelerometer. Measured noised data are shown in Figure 4(a).

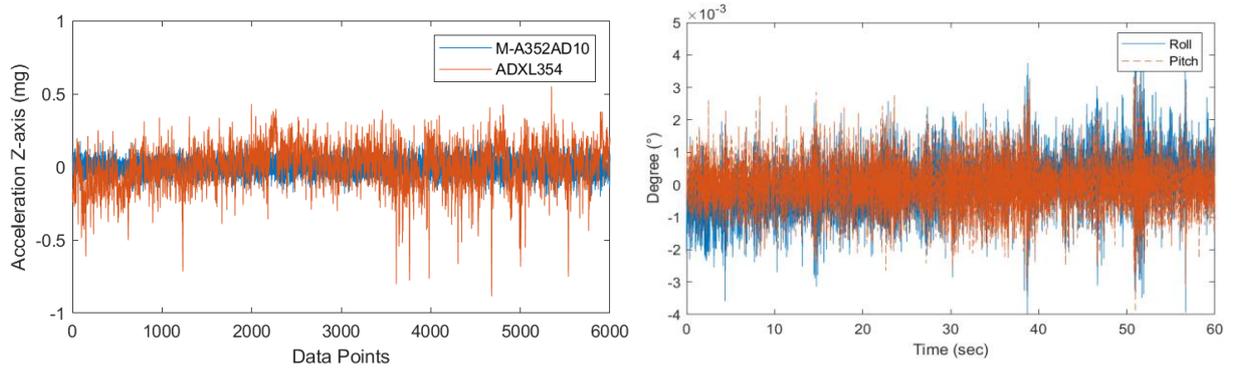

Fig. 4: (a) Acceleration noise level of ADXL354 and M-A352AD10 (b) tilt noise level of M-A352AD10

To validate the performance of tilt measurement, which is estimated from the acceleration in real-time, was progressed by using high precision rotation stage PRMTZ8/M made by THORLABS in Hanbat University, Korea. (Figure 5)

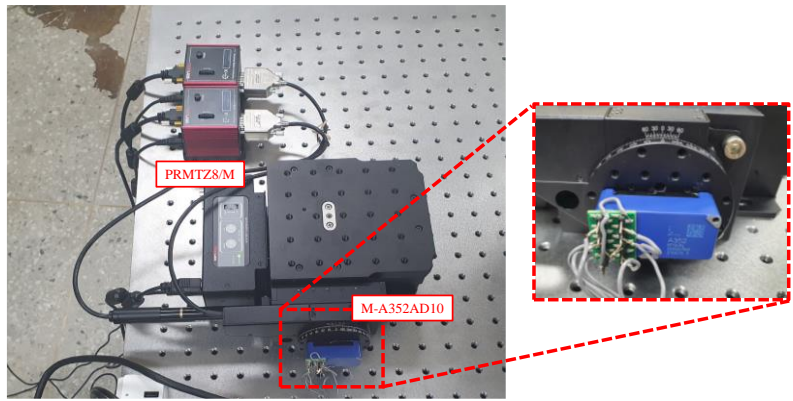

Fig. 5: Experiment setup for tilt validation

Tilt readings from the high precision rotation stage and tilt from the sensor were compared. For the accuracy validation, two cases were conducted: 0 to 25 degrees with increments of 5 degrees and micro-level tilt case. Proposed sensor showed error range within 1.4% and 1.67% respectively for both cases, which proves that the sensing system can accurately measure the tilt.

Table 1: Tilt comparison between proposed sensor and rotation stage

| PRMTZ8/M (°) | M-A352AD10 (°) | Difference (°) |
| --- | --- | --- |
| 1.000 | 0.989 | 0.011 |
| 0.100 | 0.101 | 0.011 |
| 0.010 | 0.010 | 0 |
| 0.001 | 0.001 | 0 |
| 5.000 | 4.916 | 0.084 |
| 10.000 | 9.924 | 0.076 |
| 15.000 | 14.934 | 0.066 |
| 20.000 | 19.930 | 0.070 |
| 25.000 | 24.826 | 0.174 |

Moreover, noise level of roll, pitch tilt was measured in less-vibration condition like acceleration. Figure 4(b) shows the graph of measured noise and the RMSE was evaluated to be 0.0009° and 0.0008° for roll and pitch respectively, showing high sensitivity for tilt.

### 2.1.5 Wireless Communication System

Port structure is usually hard to access due to security and safety issues. Therefore, it has been important to establish sophisticated wireless communication system that is robust to data loss. With the help of development in wireless communication technology, a modem named LM6 by M2MNET was selected, which provides state-of-the-art LTE CAT 1 communication. This communication supports upload/download speed up to 10Mbps, enabling instant data and image upload to the server. The modem was connected to RPI 3B+ through USB port. The data are transmitted in two different methods to effectively reduce the occurrence of data loss if one of them fails. First method is to send time series data in real-time. To achieve this goal, MQTT protocol is used with AT commands provided by the modem. Measured data were packaged by each 5 rows of data in a JSON format and sent to the server with prescribed topic. Another method is sending the CSV containing overall data and captured JPG files to the server directly. For this case, USB network mode was used, which makes the RPI 3B+ to be connected to the internet environment.

### 2.1.6 Thermometer and Webcam

Besides, a thermometer (SHT30) was applied to the sensing system. Temperature is also one of the important factors since it can show the relationship between structural behavior according to the temperature changes. Since applied thermometer is not waterproof, a custom 3-D printed jig was added to protect the thermometer from the water drops in outdoor condition. The thermometer was connected to RPI 3B+ through I2C communication protocol and the data was measured once per trigger event. Image acquisition from the site is another main feature required as the object detection of the ship needs to be carried out. A webcam was applied to the system to easily capture the image of the site with RPI 3B+. The used webcam supports waterproof, 5MP resolution, and embedded infrared ray for night condition that can augment accuracy of object detection.

## 2.2 Trigger System

The proposed trigger system is the way to initiate the sensing system to enter actual sensing mode, and this mechanism helps to accurately sense the exact berthing impact while avoiding accumulation of meaningless data and power waste. The trigger system consists of two parts: predefined schedule and long rangefinder based. Both systems are adequately operated to complement each other.

### 2.2.1 Predefined Schedule based

The main purpose of predefined schedule based trigger is to periodically accumulate long-term trends of the data, check the sensor status and collect images for ship detection. The trigger is set by the internal threading timer module running in the Python script. When timer is triggered, sensor directly enters to the sensing mode and the captured image is immediately sent to the cloud server for the AI-based object detection. Depending on the detection result from the server, the system decides sensing time: for example, 30 seconds and 20 minutes for no-ship condition and ship-presence condition respectively. In addition, parameters like the trigger period and sensing time can be remotely adjusted by editing server's MySQL database. Unfortunately, due to battery issue, the trigger period cannot be set to short, which means that the ship berthing event can be missed if timing is off. Therefore, long rangefinder based system has been devised to assist the weakness of predefined schedule based trigger alone itself.

#### 2.2.2 Long rangefinder based

For more accurate detection of ship berthing timing, a long rangefinder was introduced. The used long rangefinder is called LW20 by LightWare, featuring IP67 waterproof and long range measurements up to 100 meters. To obtain distance readings from the long rangefinder, external MCU called Teensy 4.0 was used in a form of distance data acquisition sensor. The independent system was suggested for distance trigger in order to lower the computation complexity of the main MCU RPI 3B+. Also, Teensy 4.0's low power consumption compared to RPI 3B+ was suitable as the long rangefinder is needed to be operated all the time. LW20 rangefinder was connected to Teensy 4.0 with I2C communication protocol and the measured data are serially logged to RPI 3B+ with sampling rate of 1Hz. In order to trigger the sensing system with long rangefinder based system, predefined distance threshold is used to predict the approach of the ship. The threshold can also be adjusted remotely through the MySQL just like predefined schedule based.

### 2.3 AI-Cloud System

#### 2.3.1 Cloud processing and DB

Amazon Elastic Cloud Compute (EC2) instance type of c5d.xlarge with Ubuntu environment was initiated. The server is composed of several parts with the help of Flask, Node.js, influxDB, PyTorch and MySQL. Flask server is used for JPG and CSV files collection from the sensor. When the server gets request via sensor's remote URL access, allocated codes are operated to save the JPG and CSV to defined directory according to its ship presence and sensor ID. Flask also does another major function when image data is received; PyTorch based Faster R-CNN inference model with trained weights is triggered for ship detection (Section 2.3.2). Node.js takes the role of processing real-time MQTT data transmission. The server's MQTT broker is based on Mosquitto, and the script was coded to subscribe all the topics generated form the sensor by using wildcard concept of MQTT. The subscriber(server) can collect all the data from multiple sensors and process them based on the subtopics. Process includes parsing data by the column and transfer them to InfluxDB, which is a database specialized for time-series data.

Data from each sensor consists of main and subordinate data. Main data such as tri-axial acceleration and tilt data are stored in "sensor#" schema (# denotes sensor ID) of InfluxDB. In case of subordinate data which includes trigger time, distance, temperature, and trigger type are stored in "sensor#_info" schema of InfluxDB. MySQL is mainly used for remote control over the sensor in the site. Sensor system is programmed to regularly check the MySQL through query in adequate condition. This enables the sensor to change the parameters and figure out the result of object detection without any intervention of human. Figure 6 describes about the flow of cloud computing algorithms in detail.

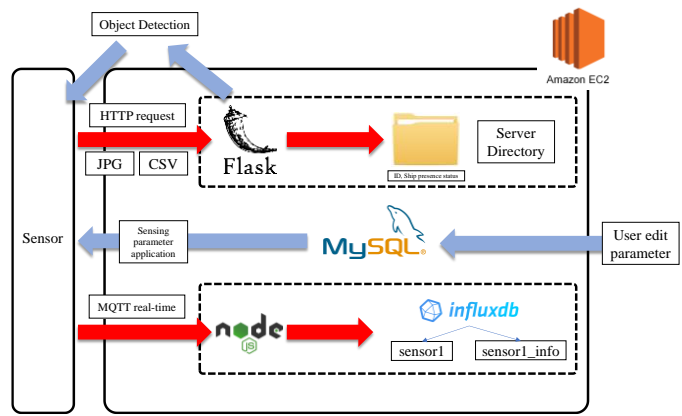

Fig. 6: Proposed AI-Cloud system flowchart

### 2.3.2 Faster R-CNN based ship detection

When trigger system triggers, the sensor needs to clarify whether ship exists or not to further determine the sensing time and correct database allocation. To do this, help of deep learning, especially object detection, has been suggested. The object detection that has been used is called Faster R-CNN, which was developed by research team in Microsoft. It has evolved from R-CNN, Fast R-CNN, and eventually could accurately and quickly predict the location and class of the object. The main contribution of Faster R-CNN is that it proposed the concept of region proposal network (RPN) and anchor boxes that could significantly reduce the computational time and make the network to be trained end-to-end.

For the model training, dataset was first established through Google web image crawling. 233 images containing ship object were fed to the network after labelling. Labelling was progressed by the Python based open-source annotation tool called Napari. The ground truth bounding box was annotated with the label of 'Ship'. 80% of the dataset was used for the training and the rest was equally separated in half to validation and test dataset respectively. With the help of transfer learning concept, the model could accurately detect the object even if the dataset was not large enough. Figure 7 shows how the Faster R-CNN network is composed to train and detect the object when image is fed to the model.

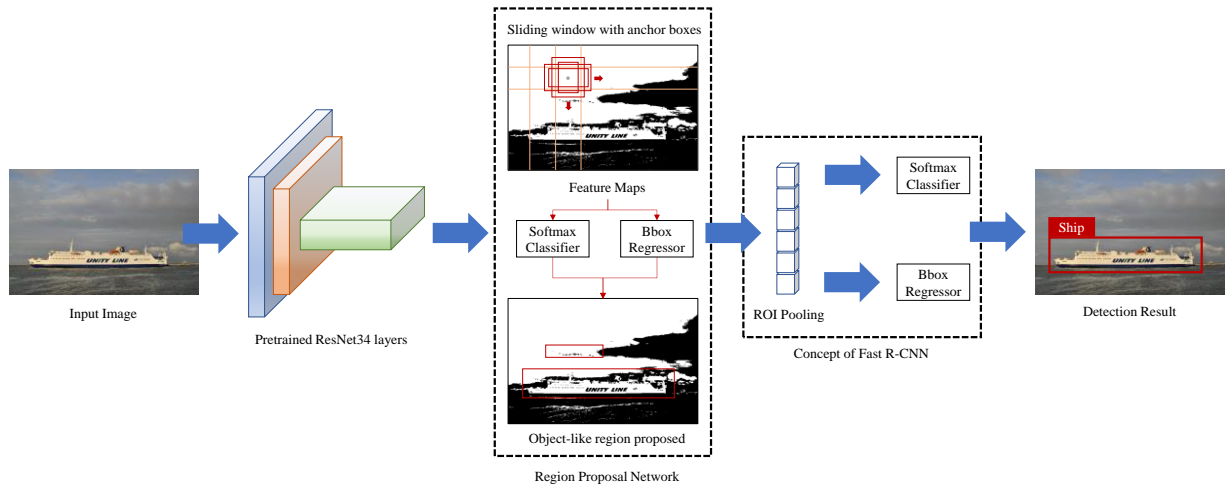

Fig. 7: Architecture of proposed Faster R-CNN model

For the convolutional network, ResNet-34 (He et al., 2016) pretrained with ImageNet dataset was used for the feature extraction. Since the backbone network is trained with the ImageNet dataset, the custom dataset needed to be transformed to adequate size. Therefore, preprocessing of changing the size of input image by 1024x1024 pixel was applied. For the image normalization, mean and standard deviation of image were selected with the values generally used in ImageNet training (Deng et al., 2009). Besides, image augmentation techniques such as horizontal and vertical flip were added in transformation stage. For the aspect ratio and size of the anchor box, 5 different sizes and 3 different ratios were adopted. In case of Intersection of Union (IoU) threshold, which is expressed on equation (1.2):

$$IoU(B_1, B_2) = \frac{B_1 \cap B_2}{B_1 \cup B_2} \qquad (1.2)$$

where $B_1, B_2$ are ground truth bounding box and predicted bounding box area respectively, value of 0.5 was used to determine true positive and false positive. Detailed information about the parameters used in training is listed on the Table 2.

Table 2: Parameters used for model training

| Parameter | Value |
| --- | --- |
| Conv network | ResNet-34 |
| Input Image Size | 1024 pixel |
| Image mean | [0.485, 0.456, 0.406] |
| Image std | [0.229, 0.224, 0.225] |
| Batch size | 2 |
| Learning rate | 0.001 |
| Maximum epochs | 500 |
| Anchor size | [32, 64, 128, 256, 512] |
| Anchor aspect ratio | [0.5, 1.0, 2.0] |
| IoU threshold | 0.5 |
| Patience | 50 |

For the training part, an early stopping technique is used to efficiently train the model and avoid overfitting issues. For each epoch, validation is processed, and average precision (AP) is calculated. If AP of validation decreases, the model is considered to be overfitted. At that point, checkpoint is saved and model gets 50 more epochs of chance (patience) for the case of further improvement. Finally, the model is stopped earlier if there is no more improvement during patience period, and the model with best performance is chosen from the generated checkpoint.

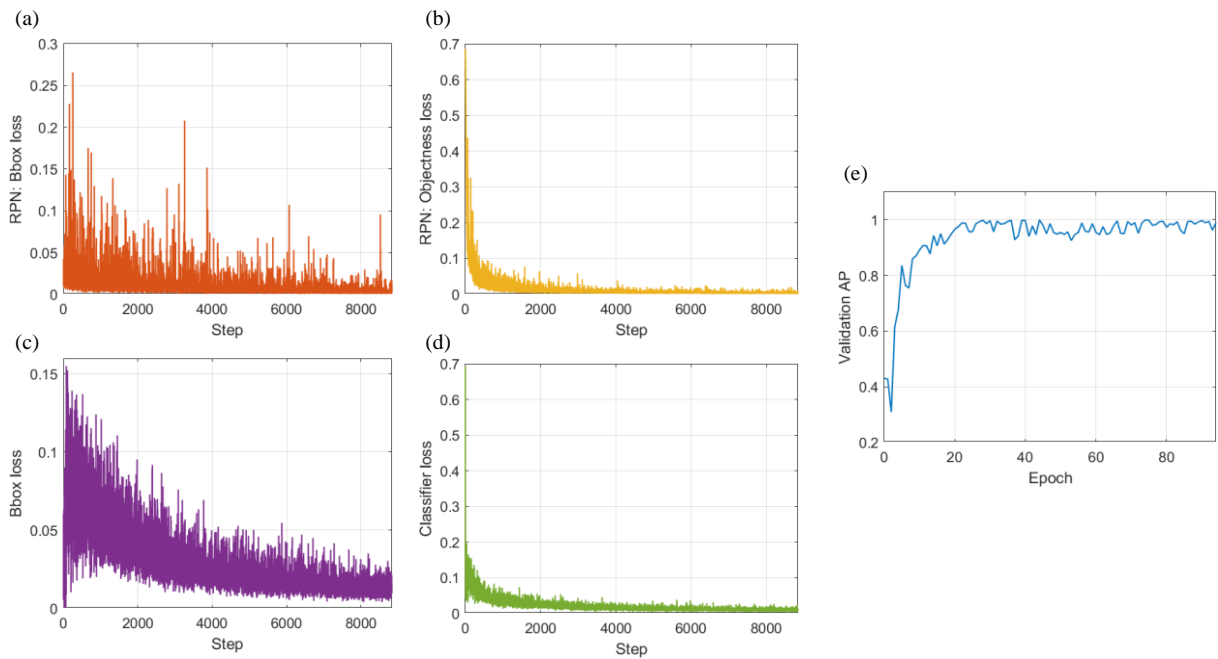

Fig. 8: Loss and AP curves during training of the model

On epoch 44, step 4184, checkpoint was created. During the patience of 50 epochs, there was no improvement of the model. Thus, the training has stopped earlier on epoch 94, not reaching to the maximum epoch which was defined to 500. The best model was selected with checkpoint model created on epoch 44. The test result for the 24 images of test dataset showed AP of 0.92, which was quite accurate. The detailed logs of loss changes during the training are shown on the figure 8.

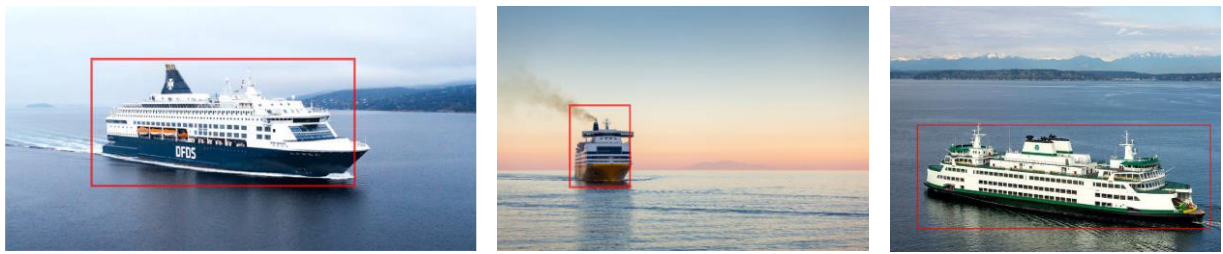

Fig. 9: Inference result of random ship image

On figure 9, inference results with the trained model are shown. Images shown on the figure 9 have never opened to the model during the training process. The inference results show that the created AI model can accurately detect the ship and figure out the location within the image through the bounding box. The model was then attached to the created AWS EC2 server as a part of proposed system. The inference time required on the server for a single image was around 0.2s, which also showed the possibility of real-time ship monitoring.

## 3. FIELD APPLICATION

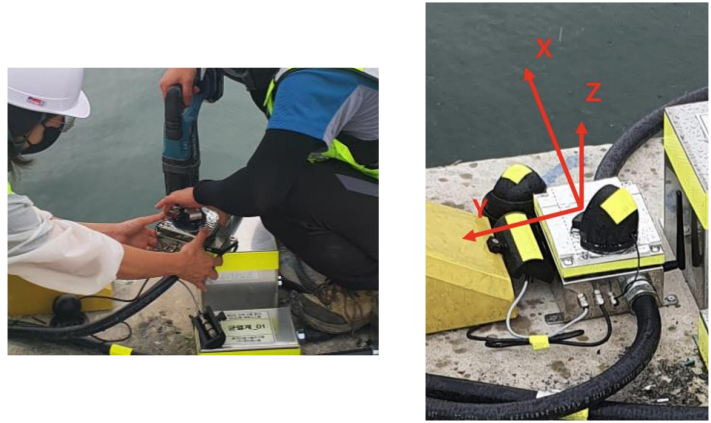

Fig. 10: Proposed sensing system installed in the field

The proposed sensing system was deployed in the XX port in Korea. (Due to security issue, the name of the port hereafter referred to as XX port and some of the images may be blurred). The sensor was deployed near to the fender system which absorbs the berthing impact and anchor bolts were used for fix to the ground. The application was delivered as a preliminary investigation to validate the applicability of the proposed system in real site condition. The system was installed on 30 May 2022 and uninstalled on 1 June 2022. Figure 10 shows the installation process and the direction of acceleration axis.

## 4. DISCUSSION

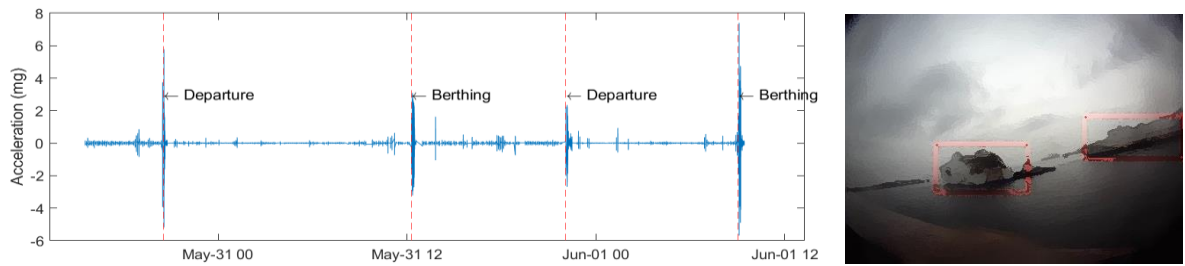

Fig. 11: (a) x-axis acceleration data (b) detection of non-berthing ship

Figure 11 shows the x-axis acceleration data collected from the field. Since the permitted time for application was short, period for schedule-based trigger was set to 5 minutes to get enough data. During the application period, predefined schedule-based trigger was successfully operated as the retrieved data on the Flask server had exact 5-minute interval when ship was not detected. The server detected 4 times of ship through long rangefinder trigger and AI object detection. The captured image and the berthing/departure schedule provided by the port authority was also 4 times. This supports that the proposed trigger system and Faster R-CNN based ship detection algorithm worked well as expected. One of the issues for the field application in aspect of ship detection was that multiple ships tends to appear within an image. There was a possibility that the proposed detection algorithm might misunderstand a standby or passing ship as a berthing event like Figure 11(b). Therefore, a new algorithm which determine the ship berthing if the coordinate and area of bounding box are within the limit was added in the beginning. Figure 12 shows improved algorithm result for 2 berthing events occurred during application period.

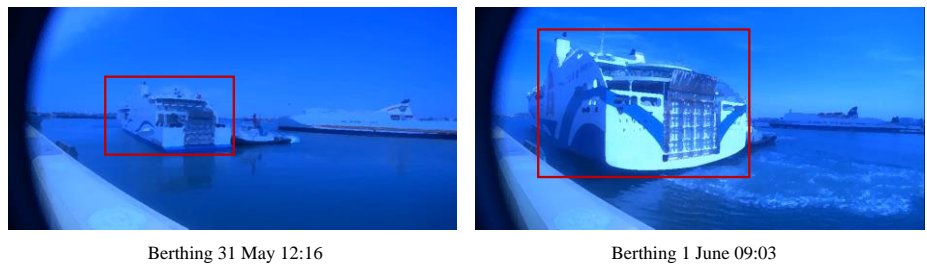

Berthing 31 May 12:16     Berthing 1 June 09:03

Fig. 12: Object detection result based on bounding box coordinate and area

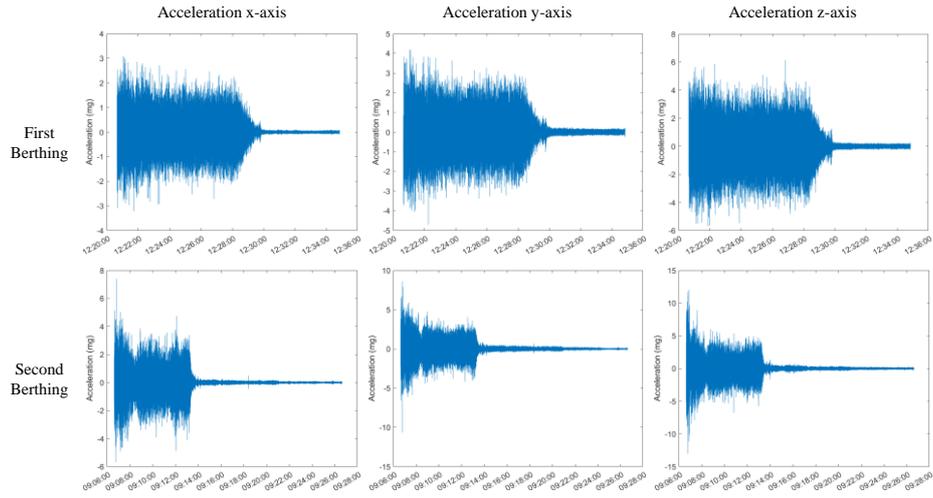

Fig. 13: Acquired acceleration data when berthing occurred

Figure 13 is the graph of berthing data acquired from the port. It describes data about two berthing events with x, y, and z acceleration. Berthing events showed higher vibration amplitude than normal condition. Second berthing event occurred on 1st June showed the highest amplitudes which are 7.398mg, 8.565mg, 12.040mg for each axis.

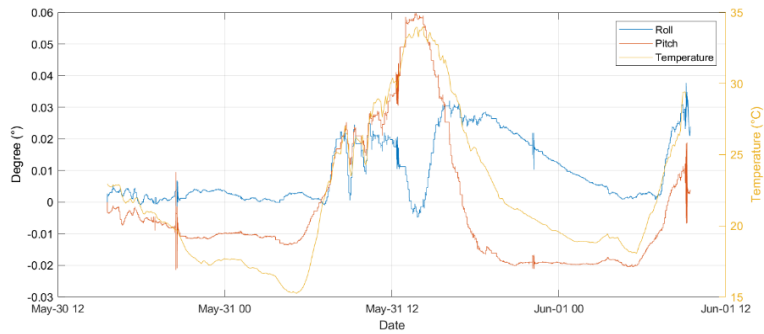

Fig. 14: Roll and pitch tilt changes during application and effect of temperature

Roll and pitch tilt data were also analyzed with the temperature of the field. For the roll tilt, the maximum change during the application was 0.0416° and 0.0798° for the pitch tilt. Pitch tends to follow the trend of temperature, while roll shows opposite behavior for the temperature higher than 25°C. The proposed system could successfully estimate the tilt from acceleration, which implies the possibility of structural analysis in various aspects. By tracking and monitoring the change of the tilt changes and maximum acceleration level, it would be possible to assess the structural condition and anomaly of the port structure.

## 5. CONCLUSION

This paper proposes an optimized overall wireless sensor system that enables structural health monitoring of port based on vibration. Vibration based structural analysis is generally used in civil engineering area, but the application is mostly hindered for port structure. Port's massive mass makes itself to vibrate with small magnitude, which is hard to be sensed by typical accelerometer. In addition, most of the sensing systems applied on the field provide constant sensing feature, consuming lots of energy and server storage with meaningless data. This feature makes the human to intervene throughout the data processing and visit the field for battery change while it is difficult due to security issue. To overcome difficulties, a novel sensing system for port structure is proposed.

Developed sensing system is composed of EPSON M-A352 accelerometer, featuring 0.2μG/√Hz of noise density, and scale factor of $2^{-24}$G/LSB. Tilt estimation based on the acceleration is also proposed to provide additional information for health monitoring without applying additional sensor. Through the lab-scale test, acceleration and tilt are proven to be appropriate for port structure application, showing high-sensitivity compared to common MEMS accelerometer. For the wireless communication, state-of-the-art LTE CAT 1 technology is applied to send data to the server with MQTT and HTTP. Trigger system is also adapted to minimize the amount of data acquisition while accurately get berthing event, which is one of the most important data for vibration-based condition assessment. Predefined schedule and long rangefinder modes are used for trigger system. For the determination of ship presence after trigger, Faster R-CNN object detection was applied to accurately detect the ship. Proposed cloud system is fully automated, which enables the data to be retrieved, analyzed, and stored systematically. Developed object detection model was trained with custom dataset and showed AP of 0.92 for the test dataset. Finally, developed system was applied to the XX port in Korea for 30 May to 1 June 2022 to validate application to the actual field. During the application period, two times of berthing event occurred, and proposed trigger and AI system could successfully detect the berthing event. Acquired berthing vibration was further analyzed, in aspects of amplitude, tilt, and temperature and checked the possibility of vibration-based port monitoring via proposed system. Future studies should investigate the certainty of long-term application, multiple sensor deployment and analysis of long-term data for actual condition assessment. In addition, establishing threshold for vibration and tilt through such as finite element analysis might prove an important area for future research to achieve digital twin of the port.